\documentclass[prl,article,twocolumn]{revtex4}
\usepackage{graphicx}
\usepackage{epsfig}
\usepackage{amsmath}
\usepackage{amsfonts}
\usepackage{amssymb}
\providecommand{\U}[1]{\protect\rule{.1in}{.1in}}
\providecommand{\U}[1]{\protect\rule{.1in}{.1in}}

\begin{document}

\title {Effect of Laughlin correlations on crystalline \\ mean field solutions of the 2DEG  
in FQHE regime }

\author{Alejandro Cabo}
\affiliation{\it Grupo de F\'{\i}sica Te\'orica, Instituto de
Cibern\'etica, Matem\'atica y F\'{\i}sica,
Calle E, No. 309, Vedado, La Habana, Cuba\\
and\\
The Abdus Salam International Centre for Theoretical Physics,
Trieste, Italy,}

\author{ F. Claro}
\affiliation{\it Facultad de F\'{\i}sica, Pontificia Universidad Cat\'{o}lica de Chile,\\
Vicu\~{n}a Mackenna 4860, Santiago, Chile\\
and\\
The Abdus Salam International Centre for Theoretical Physics,
Trieste, Italy,}

\author{  Danny Martinez-Pedrera}
\affiliation{\it Deutsches Elektronen-Synchrotron (DESY), Notkestrasse 85, D-22607, Hamburg, Germany\\
and\\
The Abdus Salam International Centre for Theoretical Physics,
Trieste, Italy.}

\begin{abstract}

The energy per particle of  many body wavefunctions that mix
Laughlin liquid with crystalline correlations for periodic samples
in the Haldane-Rezayi configuration is numerically evaluated  for
periodic samples. The Monte Carlo algorithm is employed and  the
wave functions are constructed in such a way that have the same
zeroes as the periodic Laughlin states. Results with up to 16
particles show that these  trial wavefunctions have  lower energy
than the periodic Laughlin states for finite samples even at
$\nu=\frac {1}{3}$.  Preliminary results for 36 particles suggest
that this tendency could reach the thermodynamic limit. These
results get relevance in view of the very recent experimental
measures that indicate the presence of periodic structures in the
2DEG for  extremely small temperatures and clean samples, inclusive
at main FQHE filling fractions $\nu=\frac {1}{3},\frac {2}{3} $.
\bigskip

\noindent PACS numbers: 73.43.Cd,73.43.Nq, 73.22.Gk
\end{abstract}
\maketitle

Although the Fractional Quantum Hall Effect is described in its
relevant aspects in terms of Jastrow-like many-body wave functions,
a link between this variational approach and hamiltonian solutions
is still desirable. The states introduced by Laughlin cleanly
incorporate the tendency of particles to be as far away as possible
from each other. On the other hand the Hartree-Fock states discussed
in the literature (See [\onlinecite{alamos1}] and the references
therein) are approximate solutions of the Hamiltonian and may have
valuable information not taken care of by the variational states.
With this possibility in mind we have constructed a new type of
trial wave function that incorporates a part of each approach as
described below. Precisely in these days this construction  gets
relevance thanks to very recent experiments that had detected the
presence of surprising periodic structures in the 2DEG for highly
clean samples subject to extremely small temperatures \cite{dial}.
Assuming that the proposed here wavefunctions maintain their
properties of showing lower energies than the so called periodic
Laughlin states (in the Haldane-Rezayi periodic scheme [\onlinecite{halrez}]) in the
thermodynamic limit, they could have the opportunity of describing
the detected periodic structures. Qualitatively, the energy
dependence on the sample size obtained here suggests that the energy
is lower in the limit of infinite size samples. However, this
important conclusion needs for more extensive calculations for its
full confirmation, that are expected to be considered elsewhere.

Analytic Hartree-Fock solutions in the lowest Landau level may be
written down for filling fractions of the form $\nu=1/q$. This
unusual property was discovered after numerical results showed that
there is a self-consistent charge density wave solution (CDW) to the
Hartee-Fock (HF) equations that has a zero of order $q-1$ in each
CDW plaquette. This property was then used to diagonalize the
eigenvalue matrix. The resulting charge density corresponds to a
lattice of holes, with percolating ridges that surround the zeroes.
The form of the associated single particle wavefunctions suggests a
way of constructing our new state incorporating both the Laughlin
correlation in part, as well as the HF crystalline correlations(See
[\onlinecite{alamos1}]).

The appropriate framework to achieve this on the plane using numerical
procedures is to impose periodic boundary conditions on the single particle
states, as suggested by Haldane and Rezayi many years ago
[\onlinecite{halrez}]. This is done in Ref. [\onlinecite{alamos2}]. The main
idea is to exploit the fact that the mean field Slater determinant in this
picture can be written as the product of another determinantal function
containing the whole dependence on the quantum numbers of the single particle
HF states, times a factor whose zeroes are spatially fixed and periodic with
the periodicity of the density. Therefore, the position of those zeroes has no
dependence whatever on the set of quantum numbers of the filled mean field
states, and has thus been factored out of the Slater determinant. Moreover,
the number of zeroes of those kinematical factors as a function of any of the
identical particle coordinates is just $(q-1)N_{e},$ where $N_{e}$ is the
number of electrons.

 Let us begin by writing the explicit form of the counterparts of the Laughlin wavefunction 
 in the Haldane-Rezayi scheme for implementing periodic boundary conditions
\begin{align*}
\Psi_{L}  &  =\exp(-\sum_{i=1,2...N_{e}}\frac{y_{i}^{2}}{2r_{o}^{2}})\left\{
\vartheta_{1}(\frac{\pi}{L}(Z^{\ast}-R^{\ast})|-\tau^{\ast})\right\}
^{q}\times\\
&  \prod_{\substack{i<j\\i,j=1,2...N_{e}}}\left\{  \vartheta_{1}(\frac{\pi}%
{L}(z_{i}^{\ast}-z_{j}^{\ast})|-\tau^{\ast})\right\}  ^{q},
\end{align*}
where $z_{j}^{\ast}=x_{j}-i$ $y_{j}$ , $j=1,2,....N_{e}$ and $Z^{\ast}=X-iY$
are particles and center of mass complex coordinates, respectively. $L$ is the
cell size, $R^{\ast}=n_{1}a_{1}+n_{2}a_{2}^{\ast}$ with $n_{1}$ and $n_{2}$
integers, $\tau^{\ast}=exp(-2\pi i/6)$, and $a_{1}=\sqrt{2\pi q/\sin{(2\pi
/6)}}$, $a_{2}^{\ast}=a_{1}\tau$. The functions $\vartheta_{1}(u|\tau)$ are
the odd elliptic theta functions and vanish as the first power in u as this
variable goes to zero. We notice that as one particle approaches another,
$\Psi_{L}$ vanishes as a power $q$, so that altogether it includes
$q(N_{e}-1)$ zeroes of this kind, plus $q$ generated by the center of mass
factor. We can thus replace the factors with spatially fixed zeroes in the HF
solution, by a proper Laughlin factor to obtain the same short range behavior
(a zero of order $q$) when any two particles approach each another. However,
the presence now of the determinantal function keeps the crystalline
information associated with the optimization of the mean field problem.
Therefore, the proposed states have an a priori chance of lowering the energy
per particle of the Laughlin states. 

 Then, in the Landau gauge $\mathbf{A}=-B(y,0,0)$, as it is  proper of the Haldane-Rezayi 
 scheme, the  state ansatz being proposed here  has the explicit form
\begin{widetext}
\begin{align}
\Psi &  =\Phi_{L}(z_{1}^{\ast},z_{2}^{\ast},...z_{N_{e}}^{\ast})\text{
}D[z_{1}^{\ast},z_{2}^{\ast},...z_{N_{e}}^{\ast}]\exp(-\sum_{i=1,2...N_{e}%
}\frac{y_{i}^{2}}{2r_{o}^{2}}))\nonumber\\
\Phi_{L}(z_{1}^{\ast},z_{2}^{\ast},...z_{N_{e}}^{\ast})  &  =\exp(-iQ\text{
}Z^{\ast})\left\{  \vartheta_{1}(\frac{\pi}{L}(Z^{\ast}-R^{\ast})|-\tau^{\ast
})\right\}  ^{q-1}\prod_{\substack{i<j\\i,j=1,2...N_{e}}}\left\{
\vartheta_{1}(\frac{\pi}{L}(z_{i}^{\ast}-z_{j}^{\ast})|-\tau^{\ast})\right\}
^{q-1},\nonumber\\
D(z_{1}^{\ast},z_{2}^{\ast},...z_{N_{e}}^{\ast})  &  =\text{ }Det\text{ }%
[\chi_{\mathbf{k}_{i}}^{(0)}(z_{k}^{\ast})], \label{ansatz}%
\end{align}
\end{widetext}
where $Q =-\frac{\pi}{a}(q-1),Z=\sum_{j=1}^{Ne}z_{j}$ and
$r_{0}=\sqrt{\hbar c/|eB|}$. The functions $\chi_{\mathbf{k}}^{(0)}$
have the form
\begin{align*}
\chi_{\mathbf{k}}^{(0)}(\mathbf{x})=\exp(i\varkappa\ z^{\ast}-\frac{y^{2}%
}{2r_{o}^{2}})\times &  \prod_{R}\vartheta_{1}(\frac{\pi}{L}(z^{\ast}-R^{\ast
}-C_{\mathbf{k}})|-\tau^{\ast}).\\
\end{align*}
Here $\varkappa=-\mathbf{k}.\mathbf{a}_{1}/a_{1}$ and the argument
$C_{\mathbf{k}}$ depending on the quantum number $\mathbf{k}$ is given by\
\[
C_{\mathbf{k}}=\frac{a}{2\pi}(-\mathbf{k}.\mathbf{a}_{2}-\mathbf{k}%
.\mathbf{a}_{1}\tau^{\ast})+\frac{qa\text{ }\tau^{\ast}}{2}.
\]
Our trial wave function $\Psi$ is periodic, with a slanted periodicity region
of equal sides length $L=Na_{1}$, and a slant angle of $2\pi/6$. The number of
particles in the region is $N_{e}=N^{2}$. The momenta allowed by the periodic
boundary conditions are
\[
\mathbf{k}=\frac{n_{1}}{L}\mathbf{s}_{1}+\frac{n_{2}}{L}\mathbf{s}_{2},
\]
where the reciprocal lattice unit vectors and the normal vector are defined by%

\begin{align*}
\mathbf{s}_{1}  &  =-\frac{1}{qr_{0}^{2}}\mathbf{n}\times\mathbf{a}_{2},\text{
\ \ \ }\mathbf{s}_{2}=\frac{1}{qr_{0}^{2}}\mathbf{n}\times\mathbf{a}_{1}\\
\mathbf{n}  &  =(0,0,1),\text{ \ \ }\mathbf{a}_{i}{\small \cdot}\text{
}\mathbf{s}_{j}=2\pi\delta_{ij},
\end{align*}
while the allowed values of $n_{1}$ and $n_{2}$ are%
\begin{align*}
n_{1}  &  \in\{-\frac{N}{2},-\frac{N}{2}+1,..,0,...,\frac{N}{2}-1\},\\
n_{2}  &  \in\{-\frac{N}{2},-\frac{N}{2}+1,..,0,...,\frac{N}{2}-1\}.
\end{align*}

The evaluations of the energy per particle of both states were done 
by employing the Monte-Carlo method for
samples having a number of particles $N_{e}$ equal to $4$ \ and $16$
for the case of the Laughlin states. As for the trial wavefunctions
investigated here the calculations were done for $4$, $16$ and $36$
particles. The results are illustrated in the Tables I and II \ and
in Fig. 1. \ As mentioned  before,  the parameter $\xi$ is the one
defining  the probability of admission of new configurations  as
usually is needed to do in the Monte-Carlo algorithm. The margins of
errors reported correspond to the maximum deviation from the mean
value of a set of the last 60 percent of the evaluated energies in
the Monte-Carlo iterative process.

\begin{table}
\caption{  The results of the energy per particle for the counterparts of the  Laughlin states 
 in the Haldane-Rezayi periodic  boundary conditions. The state for $N_e=4$ was evaluated two times for different values of the Monte Carlo new configuration admission coefficient $\xi$ in order to check the independence of its value. The calculations were done for $N_e=4,16$ particles}
\bigskip
\begin{tabular}
[c]{||c||l||c||}\hline\hline
$N_{e}=N^{2}$ & $\ \ \xi$ & $ \epsilon
$\\\hline\hline
$ 4$ & $0.3$ & $-0.374476\pm0.0000697292$\\\hline\hline
$ 4$ & $0.25$ & $-0.374245\pm0.0000779082$\\\hline\hline
$16$ & $0.3$ & $-0.392032\pm0.0000880513$\\\hline\hline
\end{tabular}
\end{table}

\begin{table}
\caption{The results for the energy per particle for the trial state  proposed in this work. The same  Haldane-Rezayi periodic  boundary conditions were employed. In this case the states for $N_e=4,16$ were evaluated two times each one for different values of $\xi$ to check the independence of the result on this constant. The evaluations were  done for $N_e=4,16$ and $36$ particles in this case}
\bigskip
\begin{tabular}
[c]{||c||l||c||}\hline\hline
$N_{e}=N^{2}$ & $\ \ \xi$ & $\epsilon
$\\\hline\hline
$4$ & $0.2$ & $-0.414112\pm0.0000625317$\\\hline\hline
$ 4$ & $0.25$ & $-0.414191\pm0.0000457599$\\\hline\hline
$16$ & $0.25$ & $-0.410156\pm0.00006590$\\\hline\hline
$ 16$ & $0.2$ & $-0.410222\pm0.00008659$\\\hline\hline
$36$ & $0.25$ & $-0.410943\pm0.000486141$\\\hline\hline
\end{tabular}
\end{table}

The expectation value of the many-particle Hamiltonian in the Laughlin state
was evaluated using the Monte Carlo method for samples with $N_{e}=4,16$. For
our trial wavefunctions calculations were done for $4$, $16$ and $36$
particles. Results are shown in Tables I and II. The parameter $\xi$ defines
the probability of admission of new configurations in the Monte-Carlo
algorithm. The errors reported in the third column correspond to the mean
square of the fluctuations in the Monte Carlo output after convergence was
assured.\newline
 Our results are plotted in Fig. 1. The error bars are not resolved at the
scale of the plot. 

\begin{figure}[h]
\epsfig{figure=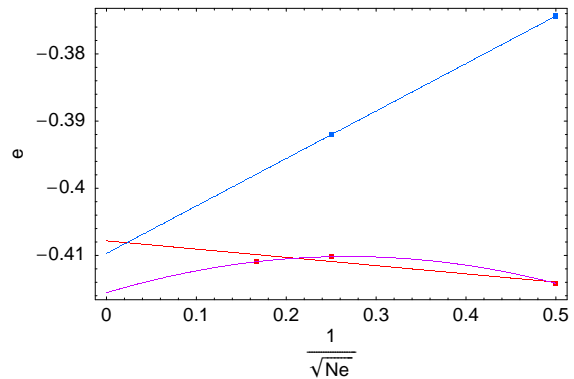,width=8cm} \vspace{0cm}
%\end{center}
\caption{Energy per particle as a function of the inverse square root of the
number of particles $N=\sqrt{N_e}$.  The picture shows that the introduction of
correlations in the HF crystalline states made their energies lower
than the ones shown by the versions of the Laughlin state
in the Haldane-Rezayi periodic scheme.   Note also that,  if the
behavior in the large $N$ limit is confirmed by more extensive
evaluations, the results will imply the existence of a ground state
with a slightly lower energy than  the Laughlin one for macroscopic
samples. The curve joining the points of evaluated energies for the new trial state 
 is a fitting of these three points to a quadratic polynomial in $1/\sqrt{N_e}$. The lower straight line
 with negative slope  is simply a linear curve of $1/\sqrt{N_e}$  minimizing the mean square deviations
 from the three measured energies.   
   }%
\label{grafico3}%
\end{figure}Note that a linear extrapolation to the thermodynamical limit
$N->\infty$ of the Laughlin state energy reproduces former estimates
of the energy of this state at $\nu=1/3$ . It is also clear that the
energy per particle of each of the computed Laughlin values lies
above that obtained for our trial wave function for any of the
evaluated finite size samples. Moreover, the behavior of the energy
of the latter for the largest number of particles evaluated,
 suggests that the energy per particle in the thermodynamic
limit is lower than the one associated with the Laughlin state.
In the contrary case,  the extrapolation curve  in  the variable $1/N$
 which join the there three evaluated points of the energy $N=2,4$ and $6$
  should change its monotonic  change of slope upon enlarging the value of $N$.  
However, being the  number of Monte Carlo method iterations  for the 36 particles state ($N=6$) 
yet limited, this indication  is not yet conclusive and further numerical evaluations will be
done to give a better foundation to this conclusion. Its validity,
clearly  leads to the idea about that the  recently detected
periodic structures in extremely perfect 2DEG at  very low
temperatures, could be associated to the here proposed translation
symmetry breaking states \cite{dial}.
 
 In ending we would like to underline that the present letter consider 
a particular HF state showing  one electron per its periodicity unit
cell. For fillings of the form $1/q$ the HF solution produces a gap
for all integers $q$. The experiment suggests, however, that even
and odd $q$ values are qualitatively different states. It has been
shown in the past that if only half electron is captured by the unit
crystalline cell this distinction is properly borne out
\cite{francisco}. Future work will extend to cover such states, and
shall be reported elsewhere.

\begin{acknowledgments}
This work was preparared  during two visits to the Abdus Salam
International Centre for Theoretical Physics, Trieste, Italy.
Support from the Condensed Matter Section is gratefully
acknowledged. One of us (A.C.) thanks the Caribbean Network on
Quantum Mechanics, Particles and Fields (Net-35) of the ICTP Office
of External Activities (OEA). F.C. acknowledges partial support
received from Fondecyt, Grants 1060650 and 7060650, and the Catholic
University of Chile.
\end{acknowledgments}
\bigskip
%\centerline{\bf References}

\end{document}